\documentclass[reprint,amsmath,amssymb,aps,prb,showpacs,floatfix,superscriptaddress]{revtex4-1} 

\usepackage[T1]{fontenc}
\usepackage{amsmath,amssymb,amsfonts}
\usepackage{txfonts}            
\usepackage{graphicx}           
\usepackage{dcolumn}            
\usepackage{hyperref}           
\hypersetup{colorlinks=true}

\def\vec#1{{\bf #1}}            
\newcommand{\hamil}{\ensuremath\mathcal{H}}
\newcommand{\sumk}{\ensuremath\sum_{k=1}^{d}}
\newcommand{\sumi}{\ensuremath\sum_{i=1}^{N}}
\newcommand{\sk}[1][]{\ensuremath s^{k}_{#1}}
\newcommand{\sx}[1][]{\ensuremath s^{x}_{#1}}
\newcommand{\sy}[1][]{\ensuremath s^{y}_{#1}}
\newcommand{\sz}[1][]{\ensuremath s^{z}_{#1}}
\newcommand{\uv}[1]{\ensuremath\vec{\hat{#1}}} 
\newcommand{\vs}[1][]{\ensuremath\vec{s}_{#1}}
\newcommand{\htri}{\ensuremath\hamil^{\text{(3D)}}}
\newcommand{\kB}{\ensuremath k_\mathrm{B}}
\newcommand{\expectation}[1]{\ensuremath\langle #1 \rangle}
 
\newcommand{\Pfrac}[3][]{\frac{\partial^{#1}#2}{\partial#3^{#1}}} 

\newcommand{\bmax}{\ensuremath \beta_{\text{max}}}
\newcommand{\cmax}{\ensuremath \chi_{\text{max}}}

\newcommand{\cdof}{\ensuremath \chi^{2}_{\text{dof}}}
\newcommand{\bmin}{\ensuremath \beta_{\text{min}}}
\newcommand{\qmin}{\ensuremath Q_{2, \text{min}}}
\newcommand{\beqh}{\ensuremath \beta_{\text{eqH}}}

\newcommand{\Pmax}{\ensuremath P_{\text{max}}}
\newcommand{\Pmin}{\ensuremath P_{\text{min}}}
\newcommand{\e}{\ensuremath\mathrm{e}}

\newcommand{\figspace}{\\\vspace{1em}}

\newcommand{\mc}[1]{\multicolumn{1}{c}{#1}} 

\DeclareMathVersion{nxbold}
\SetSymbolFont{operators}{nxbold}{OT1}{cmr} {b}{n}
\SetSymbolFont{letters}  {nxbold}{OML}{cmm} {b}{it}
\SetSymbolFont{symbols}  {nxbold}{OMS}{cmsy}{b}{n}
\makeatletter
\newcolumntype{B}[3]{>{\mathversion{nxbold}\DC@{#1}{#2}{#3}}c<{\DC@end}}
\makeatother

\newcommand{\mcb}[2]{\multicolumn{1}{B{.}{.}{#1}}{#2}} 

\begin{document}

\title{%
  First-order directional ordering transition in the three-dimensional
  compass model%
}

\author{Max H. Gerlach} \email{gerlach@thp.uni-koeln.de}
\affiliation{Institut f\"ur Theoretische Physik and Centre for
  Theoretical Sciences (NTZ), Universit\"at Leipzig, Postfach 100\,920,
  04009 Leipzig, Germany}
\affiliation{
  Institut f\"ur Theoretische Physik,
  Universit\"at zu K\"oln,
  Z\"ulpicher Str. 77,
  50937 K\"oln, Germany  
}
\author{Wolfhard Janke} \email{janke@itp.uni-leipzig.de}\homepage{http://www.physik.uni-leipzig.de/cqt. html}
\affiliation{Institut f\"ur Theoretische Physik and Centre for
  Theoretical Sciences (NTZ), Universit\"at Leipzig, Postfach 100\,920,
  04009 Leipzig, Germany}

\date{\today}

\begin{abstract}
  We study the low-temperature properties of the classical
  three-dimensional compass or $t_{2g}$ orbital model on simple-cubic
  lattices by means of comprehensive large-scale Monte Carlo
  simulations.  Our numerical results give evidence for a
  directionally ordered phase that is reached via a first-order
  transition at the temperature $T_0 = 0.098328(3) J /
  k_{\mathrm{B}}$.  To obtain our results we employ local and cluster
  update algorithms, parallel tempering and multiple histogram
  reweighting as well as model-specific screw-periodic boundary
  conditions, which help counteract severe finite-size effects.
\end{abstract}


\maketitle

\section{\label{sec:intro}Introduction}

The compass model~\cite{[{For a comprehensive recent review of the
    compass and related Kitaev models, see }] CompassKitaevReview} is
a generic model for orbital-orbital interactions in certain Mott
insulators such as various transition-metal compounds.  In systems
with partially filled orbital $3d$ shells it provides a heuristic
description for the coupling of $t_{2g}$ orbitals.  If their
interaction is dominated by the Kugel-Khomskii superexchange
mechanism, the quantum compass model is realized, while the
phonon-mediated Jahn-Teller effect gives rise to the classical compass
model.\cite{Kugel1982,Brink2004} Beyond the rich physics of orbital
order in recent years the quantum compass model has received increased
attention because it provides an alternative route to realize qubits
that are shielded from decoherence via so-called topological
protection.\cite{Doucot2005,Milman2007} In this context the model is
realized in the form of arrays of superconducting Josephson junctions,
which have already been implemented successfully in
experiments.\cite{Gladchenko2009}

While the compass model is closely related to the well-studied $O(n)$
and Heisenberg lattice spin models with nearest-neighbor interactions,
it differs from these in a fundamental aspect: It features an inherent
coupling of real space symmetry, realized by the point group of the
lattice, to the symmetry of the interactions encoded in the
Hamiltonian.  The resulting competition of exchange couplings along
the different lattice axes prevents a conventional magnetization-like
ordered phase, but still allows for long-ranged, essentially
one-dimensional directional ordering.\cite{Batista2005} The peculiar
symmetries of the compass model lead to a high degree of degeneracy in
its ground states,\cite{Nussinov2005} similarly to other orbital 
models.~\cite{CompassKitaevReview} Typically such
a degeneracy suppresses order for $T=0$, while at low, but finite
temperatures an ordered phase may still be realized through an
order-by-disorder\cite{Villain1980,Henley1989} mechanism, where certain system
configurations are favored entropically.  For both the classical and
the quantum variation of the compass model in two dimensions (2D),
earlier Monte Carlo studies have indeed established the realization of
a directionally ordered phase at low temperatures, which is reached by
a continuous thermal phase transition in the 2D Ising universality
class.\cite{Mishra2004,Wenzel2008,Wenzel2010}

Beyond that, the case of the three-dimensional (3D) compass model
remains particularly interesting as it may be significant for the
microscopic description of materials in the reach of experimental
research.  For the 3D quantum compass model high-temperature series
expansions have not shown any sign of a finite-temperature phase
transition, while the continuous transition could be confirmed for the
2D quantum compass model.\cite{Oitmaa2011}

The purpose of this paper is to shed more light on the low-temperature
properties of the compass model in three dimensions.  We present an
extensive Monte Carlo study that provides evidence for a first-order
phase transition from a high-temperature disordered phase into a
directionally ordered phase.  While simulations of the quantum model
are plagued by a negative-sign problem and hence are infeasible on
reasonably sized lattices, we can study the classical variation of the
3D compass model without prohibitive computational cost.
Nevertheless, a considerable methodological effort is required to
obtain quantitative results for two reasons: The model features very
strong finite-size effects that must be treated carefully and long
autocorrelation times near the transition point would make it hard to
collect sufficient statistics with only a naive Monte Carlo sampling
scheme.

The main part of this work is organized as follows: In
Sec.~\ref{sec:model} we formally introduce the model and discuss
some of its properties.  Section~\ref{sec:methods} describes the
setup of the simulations and the specific numerical methods employed.
Our results are presented and analyzed in Sec.~\ref{sec:results}.  
We close in Sec.~\ref{sec:summary} with conclusions and an outlook.

\section{\label{sec:model}The model}

In $d$ spatial dimensions the compass model is defined on a
simple-hypercubic lattice of size $N=L^d$ by the Hamiltonian
\begin{align}
  \label{eq:1}
  \hamil = -\sumk \sumi J_k \sk[i] \sk[i+\uv{k}].
\end{align}
Here $\sk[i]$ is the $k$-th component of a spin $\vs[i]$ at lattice
site $i$.  $J_k$ is a coupling constant depending on the lattice
direction $k$.  The nearest neighbor of site $i$ in the $k$-th
direction is indicated by $i+\uv{k}$.  In the classical compass model
the constituent spins are represented by vectors on the unit
hypersphere in $d$-dimensional space: $\vs[i] \in S^{d-1}$.  Two spins
on sites neighboring in direction $k$ only interact in their $k$-th
components.  Note that Eq.~(\ref{eq:1}) could be separated into $d$
independent one-dimensional Hamiltonians, if the directions were not
coupled by the constraint $|\vs[i]|=1$.

In this paper we limit the discussion to equal coupling constants in
every direction: $J_k \equiv J$.  The Hamiltonian of the
three-dimensional model on a cubic lattice of size $N=L^3$ then reads
\begin{align}
  \label{eq:2}
  \htri = -J \sumi \left[ \sx[i]\sx[i+\uv{x}] + \sy[i]\sy[i+\uv{y}] +
    \sz[i]\sz[i+\uv{z}] \right],
\end{align}
where the spins $\vs[i] \in S^2$ can be parametrized by azimuthal and
polar angles $\theta_i \in [0, \pi]$ and $\varphi_i \in [0,2\pi)$:
\begin{align}
  \label{eq:3}
  \vs[i] = \vec{s}(\theta_i,\varphi_i) =
  \begin{pmatrix}
    \sx[i] \\ \sy[i] \\ \sz[i]
  \end{pmatrix}
  =
  \begin{pmatrix}
    \sin \theta_i \cos \varphi_i \\
    \sin \theta_i \sin \varphi_i  \\
    \cos \theta_i
  \end{pmatrix} .
\end{align}
In this work we choose a coupling constant of $J > 0$ corresponding to
ferromagnetic interactions.

The classical compass model is obtained by taking the limit of large
spin $S$ of the quantum mechanical compass model, where the spins
would be represented by $S=1/2$ operators $\vs[i] = \frac{\hbar}{2}
(\sigma_x, \sigma_y, \sigma_z)$ with the Pauli matrices $\sigma_k$.

\begin{figure}[ht]
  \includegraphics[]{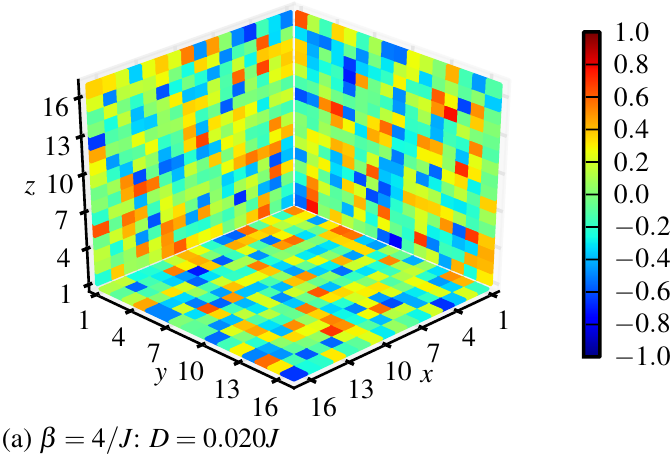}
  \figspace
  \includegraphics[]{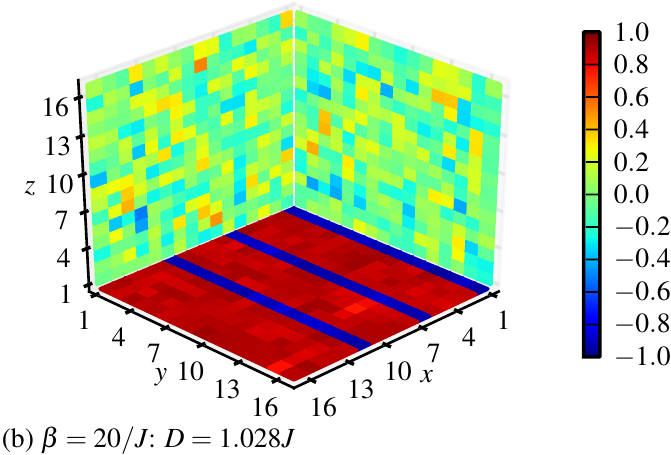}
  \caption{\label{fig:configs} (Color online) Shown are two typical
    example spin configurations of the $L=16$ system from (a) the
    disordered high-temperature phase and (b) the directionally
    ordered low-temperature phase.  On each face of the cube the
    averaged projection to the orthogonal direction of all spins at
    sites in one column above that face is given color-coded.  While
    in the high-temperature snapshot at $\beta=4/J$ no order can be
    recognized, there is a strong tendency towards linear alignment of
    the spins in the $\pm \uv{z}$-directions in the low-temperature
    snapshot at $\beta=20/J$.  }
\end{figure}

The compass model in Eq.~\eqref{eq:1} has a high number of ground
states.  To begin with, any constant spin configuration is a ground
state.  Beyond that, the model exhibits a number of discrete
symmetries, which lead to a macroscopic degeneracy of every energetic
state, including but not limited to the ground
state.\cite{Nussinov2005,CompassKitaevReview} Most importantly for
$d=3$ with open or periodic boundary conditions, Eq.~\eqref{eq:2} is
invariant under a reflection of all spins on any line of sites
parallel to one of the lattice axes across the orthogonal plane, which
leads to a $2^{3 L^2}$-fold degeneracy.  As a consequence of these
gauge-like symmetries conventional magnetic order is prohibited at any
temperature:\cite{Batista2005} $\expectation{m} =
\expectation{|\frac{1}{N} \sum_i \vs[i]|} \equiv 0$.  However,
quantities such as $\expectation{\sk[i]\sk[i+\uv{k}]}$ are invariant
under these symmetries and a special type of directional or
``nematic'' ordering is not precluded.  One can construct order
parameters that measure directional ordering characterized by
long-rang correlations in the direction of fluctuations in spin and
lattice spaces, even though magnetic ordering is absent.  This type of
order is realized by linear spin alignment parallel to the lattice
axes so that nearest-neighbor bonds carrying the lowest energy are
oriented mostly along one specific direction as illustrated in
Fig.~\ref{fig:configs}. It is not obvious to which degree the
ground-state degeneracy translates into the number of distinct
directionally ordered phases at low finite temperature.

\section{\label{sec:methods}Numerical methods}
\subsection{\label{sec:observables}Observables}

We now turn to our numerical simulations of Eq.~\eqref{eq:2} carried
out at various inverse temperatures $\beta = 1 / k_{\mathrm{B}} T$ and 
first discuss the quantities we measure.  
By $E_k = -J \sumi \sk[i] \sk[i + \uv{k}]$ with $k=x,y,z$ we denote 
the total bond energy along the $k$-th lattice axis.  Our basic 
observable is then the total energy
\begin{align}
  \label{eq:4}
  E &= E_x + E_y + E_z
\end{align}
with the corresponding heat capacity
\begin{align}
  \label{eq:5}
  C &= \Pfrac{E}{T} = \kB \beta^2 \left[ \expectation{E^2} -
    \expectation{E}^2 \right].
\end{align}
In previous studies an order parameter for directional ordering in the
two-dimensional model has been defined by the energy excess in one of
the lattice directions compared to the other
direction.\cite{Mishra2004,Wenzel2008,Wenzel2010}  Here we consider a
three-dimensional extension
\begin{align}
  \label{eq:6}
  D &= \frac{1}{N} \sqrt{(E_y-E_x)^2 + (E_z-E_y)^2 + (E_x-E_z)^2}.
\end{align}
To help with the analysis of the directional ordering phase transition
and its finite-size scaling we also consider quantities derived from
$D$: the susceptibility $\chi$ and the Binder parameter $Q_2$, which
are defined as
\begin{align}
  \label{eq:7}
  &\chi = N \left[ \expectation{D^2} - \expectation{D}^2 \right], &Q_2
  = 1 - \frac{1}{3} \frac{\expectation{D^4}}{\expectation{D^2}^2}.
\end{align}

\subsection{\label{sec:screw-peri-bound}Screw-periodic boundary
  conditions}
In most cases simulations of statistical models are carried out on
finite lattices with the topology of a torus, i.e., with periodic
boundary conditions.  The assumption is that compared to open or fixed
boundary conditions this choice minimizes finite-size surface effects,
which become irrelevant in the thermodynamic limit.

In previous studies of the two-dimensional classical compass model,
however, periodic boundary conditions have not turned out to be an
ideal choice.  In the directionally ordered low-temperature phase the
spins form essentially one-dimensional chains with decoupled rows and
columns of spins on the square lattice.  With periodic boundary
conditions the spins tend to form closed aligned loops along the
boundaries of a finite lattice.  Such excitations are particularly
stable against thermal fluctuations.  In their studies Mishra et
al. have noticed such an effect spoiling the finite-size scaling with
periodic boundary conditions\cite{Mishra2004} and suggested that the
reason may lie in the existence of a one-dimensional magnetic
correlation length $\xi_{\text{1D}}$ which exceeds the linear system
size $L$ at low temperatures.  Wenzel et al. have confirmed this
claim.\cite{Wenzel2010}

As a solution the authors of Ref.~\onlinecite{Mishra2004} have adopted
special fluctuating or annealed boundary conditions.  Here the signs
of the coupling constants on the bonds at the lattice boundaries are
allowed to fluctuate thermally.  In this way, one-dimensional chains
are effectively broken up.  While one can assume that the influence of
these $d L^{d-1}$ fluctuating bonds becomes unimportant in the
thermodynamic limit as $N=L^d\rightarrow\infty$, this choice still
constitutes a considerable modification of the model and no good
finite-size scaling theory is available for this type of boundary
conditions.

As an alternative the authors of Ref.~\onlinecite{Wenzel2010} have
proposed screw-periodic boundary conditions, which are a particular
deformation of the torus topology of regular periodic boundary
conditions.  We generalize their definition to three dimensions to
obtain boundary conditions that interconnect lines of spins along any
of the principal lattice directions.  Explicitly, the nearest
neighbors of a site $i=(x,y,z)$ in directions $\uv{x}$, $\uv{y}$,
$\uv{z}$ are specified as follows:
\begin{align}
  \label{eq:8}
  (x,y,z) + \uv{x} &=
  \begin{cases}
    (x+1, y, z),            &\text{if } x < L-1, \\
    (0, y, [z+S]\bmod L), &\text{if } x = L-1,
  \end{cases} \notag\displaybreak[1]\\
  (x,y,z) + \uv{y} &=
  \begin{cases}
    (x, y+1, z),            &\text{if } y < L-1, \\
    ([x+S]\bmod L, 0, z), &\text{if } y = L-1,
  \end{cases} \displaybreak[1]\\
  (x,y,z) + \uv{z} &=
  \begin{cases}
    (x, y, z+1),            &\text{if } z < L-1, \\
    (x, [y+S]\bmod L, 0), &\text{if } z = L-1.
  \end{cases} \notag
\end{align}
Here the screw length $S$ is a parameter that can be varied.  If $S$
is taken as one of the distinct divisors of $L$, each plane of the
lattice can be subdivided into $S$ groups of sites or ``loops'' in
each in-plane direction $\uv{k}$, which are linked as pairs of
neighbors along that direction.  With $S=0$ or $S=L$ regular periodic
boundary conditions are recovered.  With $S=1$ there are only single
loops for each direction in a plane.  The power of screw-periodic
boundary conditions lies in the fact that with a sufficiently low
choice of $S$, the loop length exceeds the magnetic correlation length
$\xi_{\text{1D}}$ already for small $L$.  Hence, linearly aligned
excitations are broken up more easily than with regular periodic
boundary conditions.  Besides that the screw-periodic boundary
conditions reduce the number of discrete symmetries in the compass
model and the energetic degeneracy of its configurations such that the
leading degeneracy factor mentioned at the end of
Sec.~\ref{sec:model} is lowered from $2^{3 L^2}$ to $2^{3 L}$.

We have found that also for the three-dimensional model regular
periodic boundary conditions lead to poor finite-size scaling results.
Moreover, the simple definition~(\ref{eq:6}) of the order parameter
$D$ is disadvantageous with these boundary conditions because it
assigns different values to configurations which differ by planar
rotations, but really show an equal degree of order.  To remedy both
problems we use screw-periodic boundary conditions according to the
definition~(\ref{eq:8}) with a choice of $S=1$.

The choice of these boundary conditions is not expected to have an
influence on the thermodynamic limit.  They have also been
successfully applied for other purposes, e.g., for the controlled
formation of tilted interfaces between ordered domains in the Ising
model.\cite{BittnerTiltedInterfaces}

\subsection{\label{sec:monte-carlo-methods}Monte Carlo methods}

In the following section we outline the Monte Carlo algorithms applied
in our simulations.

Fundamentally we use the standard Metropolis
algorithm\cite{Metropolis1953} for local single-spin updates.  In one
lattice sweep new orientations are proposed in sequential order for
the spins at all sites.  The direction of the new spin vector is
chosen randomly from a uniform distribution over the surface area of a
spherical cap centered around the original vector. To ensure proper
uniform sampling of the angular variables the spherical measure of
integration $\sin\theta\, d\theta\, d\varphi$ is respected.  During
thermalization we adjust the opening angle of this spherical cap in
such a way that an average acceptance ratio of $50\%$ is realized at
each temperature.

To reduce autocorrelation times we additionally use the
one-dimensional version of the Wolff cluster update\cite{Wolff1989}
introduced earlier for the 2D compass model\cite{Wenzel2010} in a
direct extension to the 3D model.  This update exploits one of the
discrete symmetries of the Hamiltonian, which is left invariant if a
line of neighboring spins along one of the lattice directions is
reflected about the plane orthogonal to that direction.  To construct
a cluster first a random starting site $i$ and a lattice direction
$\uv{k} \in \{\uv{x},\uv{y},\uv{z}\}$ are chosen and the spin $s_i^k
\rightarrow -s_i^k$ is flipped, then neighboring sites in directions
$\pm \uv{k}$ are added to the cluster with probability
\begin{align}
  \label{eq:9}
  P_{i,i\pm\uv{k}}(\vs[i], \vs[i\pm\uv{k}])
   = 1 - \exp\left(\min \left\{ 0, 2 \beta J \sk[i]
  \sk[i\pm\uv{k}] \right\}\right).
\end{align}
This step is iterated with the newly adjoined site $i \pm \bf{\hat{k}}$
taking the place of $i$ until no further sites are added.  All spins
in the strictly one-dimensional cluster constructed in this way are
thus flipped at the same time.  Due to the restricted set of possible
reflection planes, this update is not ergodic on its own, but must be
used in combination with local spin updates.  In our simulations $3L$
cluster updates in randomly chosen directions are followed by $N=L^3$
local updates and we count this combination as one Monte Carlo sweep.

To further reduce autocorrelation times and improve statistics we
combine these canonical algorithms with a parallel-tempering
scheme.\cite{Geyer1991,HukushimaNemoto} Different replicas of the
system are simulated simultaneously at various inverse temperatures
$\beta_k$.  We propose exchanges of system configurations between
replicas at adjunct temperature points every $100$ sweeps.  The range
of simulation temperatures is chosen according to the scheme of
constant entropy increase,\cite{Sabo2008} which clusters the
temperature points close to a phase transition and thus eases
diffusion in temperature space, which has been valuable for the
simulations on large lattices.

From the measurements taken in the various replicas we obtain time
series of the observables $D$ and $E$ at various discrete inverse
temperatures $\beta_k$.  Making use of multiple histogram reweighting
techniques\cite{Ferrenberg1989} these observables as well as the
derived quantities $\chi$, $Q_2$ and $C$ can be estimated also at
arbitrary intermediate temperatures from the optimally combined
simulation data.  We limit discretization errors by computing
per-sample weighting factors from the density of states and
reweighting observable time series directly.\cite{Chodera2007} By
applying Brent's algorithm for minimization\cite{Brent1973} we can
precisely determine extremal temperature locations and values of
$\chi$, $Q_2$ and $C$ or other quantities which are useful to
characterize the finite-size scaling behavior at a phase transition.
Estimates of the statistical uncertainties of these quantities are
obtained by performing this procedure on jackknife resampled data
sets.\cite{Efron1982,Janke2008}

\section{\label{sec:results}Results}

\begin{figure}[htp!]
  \includegraphics[]{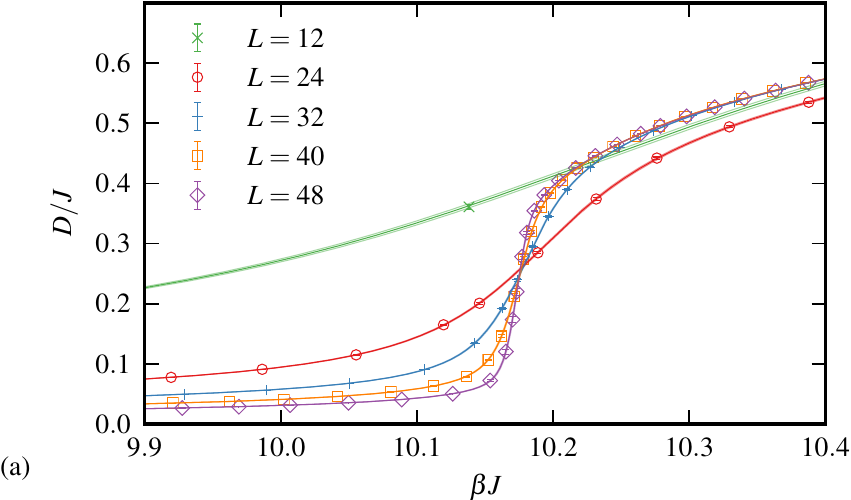}
  \figspace
  \includegraphics[]{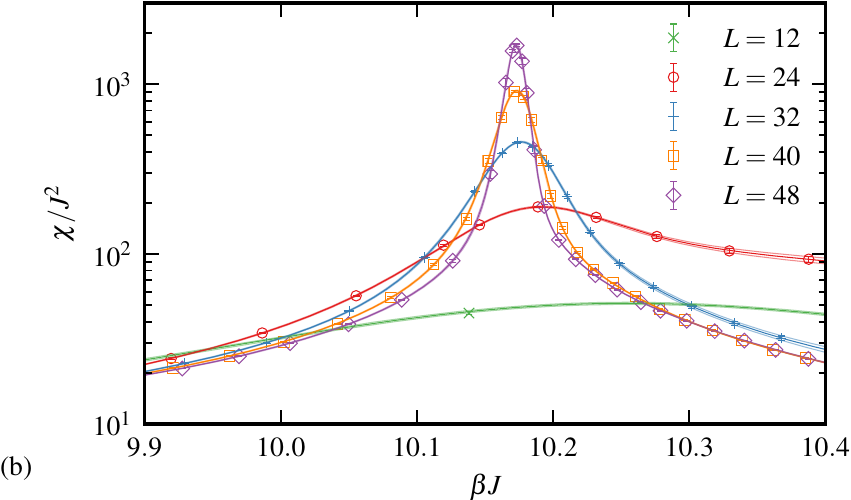}
  \figspace
  \includegraphics[]{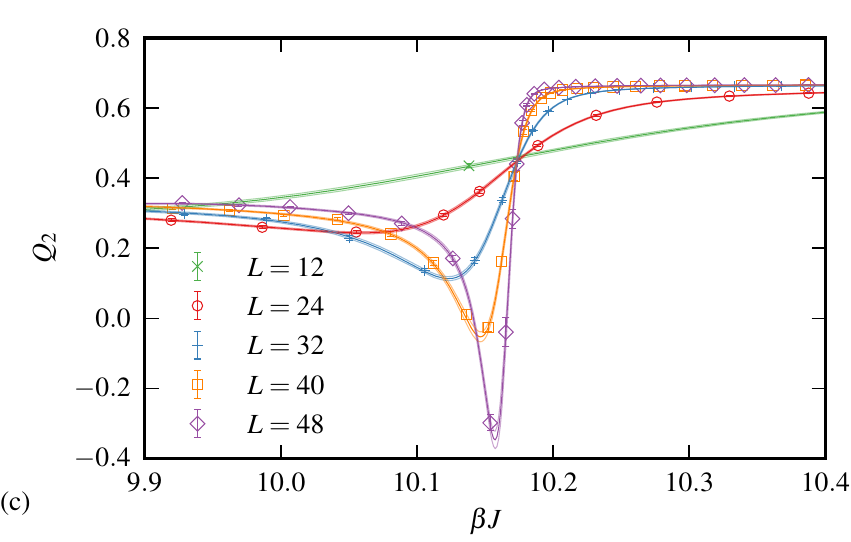}
  \caption{\label{fig:dplots} (Color online) Monte Carlo data for (a)
    the order parameter $D$, (b) its susceptibility $\chi$ and (c) the
    Binder parameter $Q_2$.  For clarity the inverse temperature range
    is limited to a region around the transition point and only
    selected lattice sizes are included in the plots.  Markers with
    error bars are estimates from single-temperature time series.
    Continuous lines are from the multiple histogram analysis with
    faint surrounding lines indicating the $1\sigma$-margin of
    statistical uncertainty. }
\end{figure}

We now present the results we obtained in our Monte Carlo simulations
that employ the methods presented in the previous section.  The 3D
compass model was simulated with screw-periodic boundary conditions
with $S=1$ on simple-cubic lattices of sizes $N=L^3$ with $L \in$
$\{8$, $12$, $16$, $20$, $24$, $28$, $32$, $36$, $40$, $44$, $48\}$.
In each case from $32$ to $64$ replicas were used in the
parallel-tempering scheme.  For the smallest lattice inverse
temperatures $\beta J$ range in $\{4,\dotsc,20\}$, while for the
largest lattice $\beta J$ was chosen from $\{9.5,\dotsc,11.5\}$.
Simulations were performed for at least some $10^7$ and up to $3.8
\times 10^7$ Monte Carlo sweeps on the largest lattice after an
equilibration phase, typically one-tenth of that length.

For all lattice sizes we observe clear indications of a thermal phase
transition around $\beta J \approx 10$ in the behavior of the order
parameter $D$, which approaches zero in the high-temperature regime
(low $\beta$) and a finite value $D>0$, which characterizes
directional ordering, at low temperatures.  The two phases are
visualized in Fig.~\ref{fig:configs}.  Note that up to thermal
fluctuations we find all spins in the ordered finite-temperature phase
to be aligned with some of the lattice axes even though the ground
states of the compass model are not restricted to have such an
orientation. Apparently fluctuations around these coaxial
configurations are favored through an order-by-disorder mechanism.

In this model with ferromagnetic couplings all spins in one aligned
row of an ordered configuration point in the same direction.  While
the scalar order parameter $D$ describes the degree of this
directional ordering and serves to clearly distinguish the phases and
identify the transition point, it does not characterize the patterns
these rows form in the ordered phase.  In this respect it would be
interesting to investigate alternative order-parameter definitions
discussed in the literature.\cite{CompassKitaevReview, Nussinov2005,
  Mishra2004} Figure~\ref{fig:configs}(b), e.g., shows the formation
of a stripe pattern, which, however, is purely an effect of our choice
of screw-periodic boundary conditions: All those spins lying in one
and the same interconnected loop are forced to point in the same
direction.  A different choice of the screw parameter $S$ would lead
to a different stripe pattern.  With periodic boundary conditions
directional ordering persists, but the aligned rows will no longer
form these visual patterns.  It is important to stress that these
differences are mere finite-size effects and become meaningless in the
thermodynamic limit.  Therefore, to clearly characterize the phase
transition a careful scaling analysis as presented below in
Sec.~\ref{sec:transition-point} is very important.

\begin{figure}[t]
  \includegraphics[]{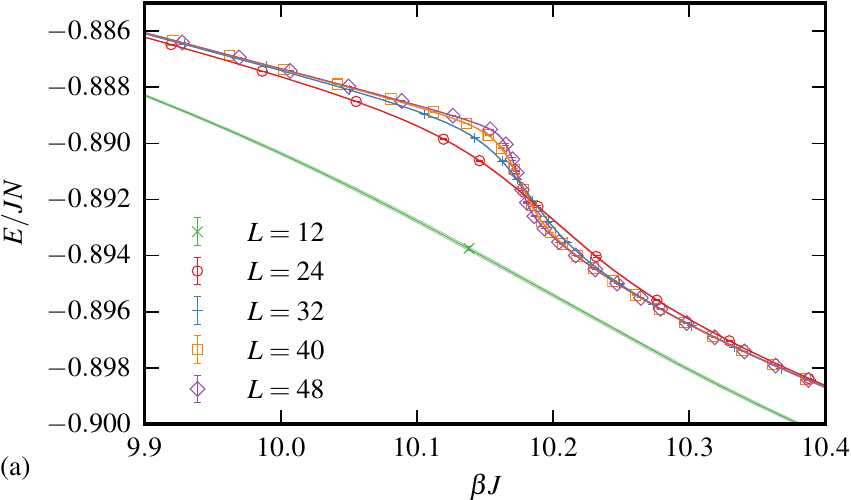}
  \figspace
  \includegraphics[]{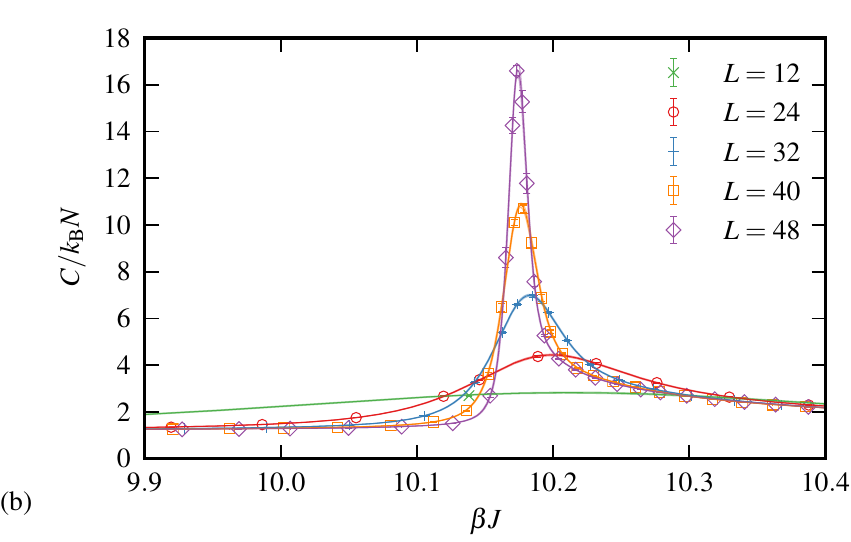}
  \caption{\label{fig:eplots} (Color online)  Monte Carlo data for
    (a) the energy per site $E/N$ and (b) the specific heat capacity
    $C/N$.  For clarity the inverse temperature range is limited to a
    region around the transition point and only selected lattice sizes
    are included in the plots.  Markers with error bars are estimates
    from single-temperature time series.  Continuous lines are from
    the multiple histogram analysis with faint surrounding lines
    indicating the $1\sigma$-margin of statistical uncertainty.  }
\end{figure}

The smoothed jump of the order parameter curve $D(\beta)$ in the
temperature region close to the transition point on different lattice
sizes can be seen in Fig.~\ref{fig:dplots}(a).  The transition is
accompanied by peaks of the susceptibility $\chi$ in
Fig.~\ref{fig:dplots}(b) and minima of the Binder parameter $Q_2$ in
Fig.~\ref{fig:dplots}(c).  On the larger lattices also bends in the
curves of the normalized energy $E(\beta)/N$ can be seen in the same
temperature region in Fig.~\ref{fig:eplots}(a) together with peaks of
the specific heat capacity $C(\beta)/N$ in Fig.~\ref{fig:eplots}(b).

\begin{figure*}[ht]
  \includegraphics[]{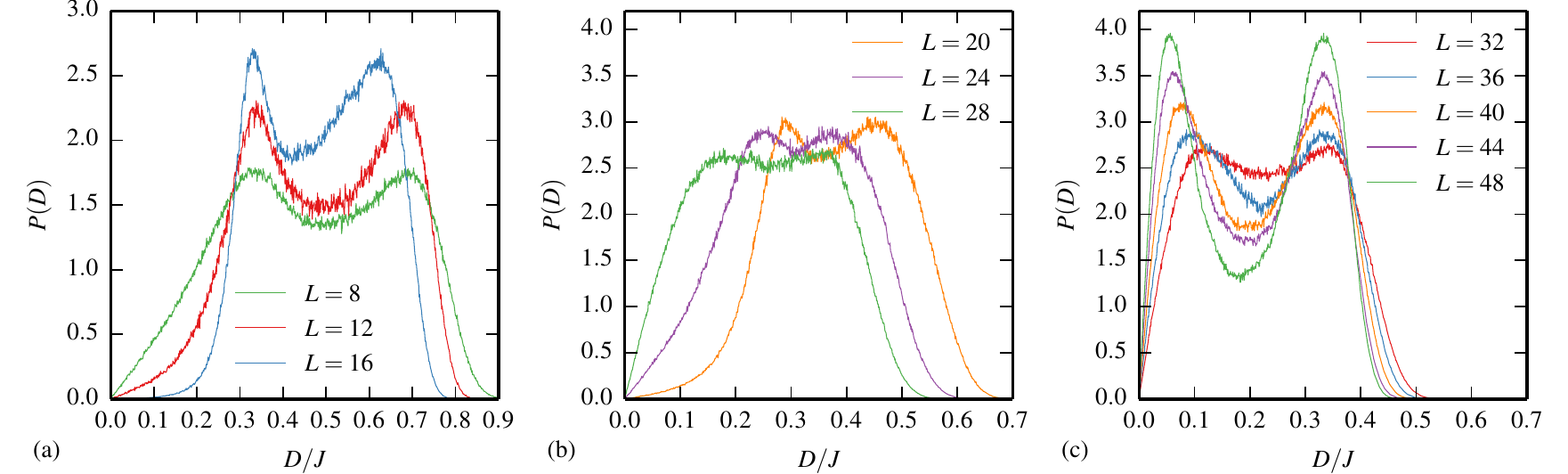}
  \caption{\label{fig:histeqheightd} (Color online)  Histograms of
    the order parameter $D$ for different lattice sizes $L$ at the
    inverse temperatures $\beqh^D(L)$, where a double-peak structure
    with equal peak height is obtained.  (a) The two peaks hinting at
    phase coexistence can be made out clearly for small lattices. (b)
    For medium sized lattices with $L<32$ the central dip shrinks with
    growing $L$. (c) For $L\ge32$ the suppression between the peaks
    grows with growing $L$.  }
\end{figure*}

\begin{figure}[bt]
  \includegraphics[]{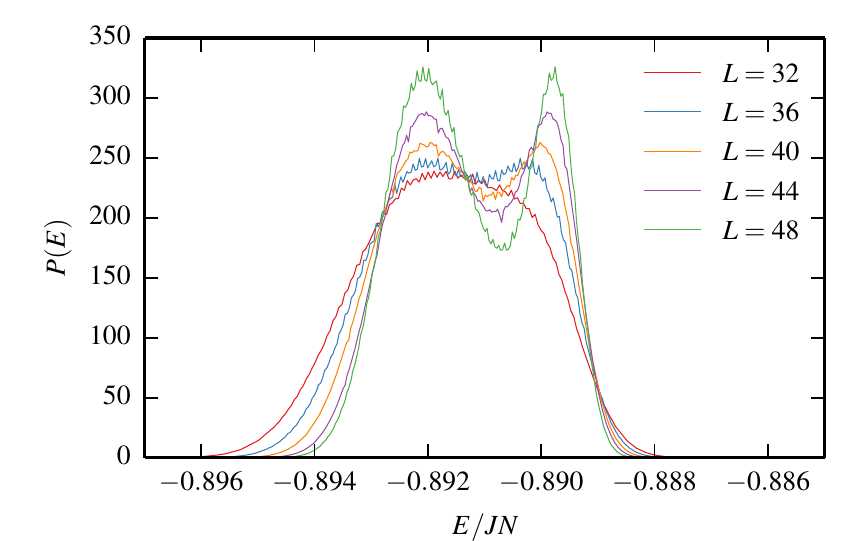}
  \caption{\label{fig:histeqheighte} (Color online)  Histograms of
    the energy per site $E/N$ for various lattice sizes $L$ at the
    inverse temperatures $\beqh^E(L)$, where a double-peak structure
    with equal peak height is obtained for $L>32$.  For $L=32$ and
    smaller lattices no double-peak distribution can be found at any
    temperature.  The $L=32$ histogram in the plot is shown only for
    comparison and is taken at a temperature close to that of the
    others.}
\end{figure}

Close to the transition we furthermore find signs for phase
coexistence, which is realized in histograms of the order parameter
$D$ with two peaks: one corresponding to a more disordered and one to
a more ordered phase.  By combining our reweighting and optimization
algorithms, we can precisely estimate the inverse temperatures
$\beqh^D(L)$, where the two peaks of the probability density $P(D)$
have equal height.  The estimates for $P(D)$ at all lattice sizes are
shown in Fig.~\ref{fig:histeqheightd}.  The double-peak structure is
already present in the smallest system studied here with $L=8$, but
from $L=16$ to $L=28$ the relative suppression at the center of the
probability distributions successively goes down and up to $L=24$ the
two peaks move closer together.  Then, starting from $L=32$, the
behavior changes again: The dip between the two peaks grows with $L$
and also their separation no longer shrinks.  Moreover, from $L=36$ on
there are also double-peak structures in the histograms of the energy
$E$.  See Fig.~\ref{fig:histeqheighte} for the distributions $P(E)$
measured at the corresponding inverse temperatures $\beqh^E(L)$.

Table~\ref{tab:pseudotemps} lists the estimated values of
$\bmax^{\chi}(L)$, $\cmax(L)$, $\bmax^C(L)$, $C_{\text{max}}(L) / N$,
$\bmin^{Q_2}(L)$, $\qmin(L)$, $\beqh^D(L)$ and $\beqh^E(L)$ for all
studied lattice sizes $L$.  The signs for phase-coexistence at the
transition temperature and the minima of the Binder parameter hint at
a first-order phase transition in the thermodynamic limit.  In the
following we study finite-size scaling relations for the measured
quantities to further support or rebut this claim.  Even with the
application of special screw-periodic boundary conditions finite-size
effects appear to be rather severe with an irregular behavior for $L
\le 32$.

\begin{table*}[htbp!]
  \caption{\label{tab:pseudotemps} Lattice-size dependent inverse pseudotransition
    temperatures.  Listed are the inverse temperature locations
    $\bmax^{\chi}(L)$, $\bmax^C(L)$ and $\bmin^{Q_2}(L)$ of the
    extrema of the susceptibility, specific heat and Binder parameter
    together with the extreme values $\cmax(L)$, $C_{\text{max}}(L) / N$ and
    $\qmin(L)$ as well as the inverse temperatures $\beqh^D(L)$ and
    $\beqh^E(L)$ where the histograms of the order parameter $D$ or
    the energy $E$ have two peaks of equal height together with the ratios
    of the estimated probabilities $\Pmax(L)/\Pmin(L)$ at the highest peak and at the
    lowest point in the dip.  }
  \begin{ruledtabular}
    \begin{tabular}{
      r
      D{.}{.}{2.6}
      r
      D{.}{.}{2.6}
      D{.}{.}{2.5}
      D{.}{.}{2.5}
      D{.}{.}{3.5}
      D{.}{.}{2.6}
      D{.}{.}{1.4}
      D{.}{.}{2.6}
      D{.}{.}{1.4}
      }
      \mc{$L$} & \mc{$\bmax^{\chi}J$} & \mc{$\cmax / J^2$} &
                                                             \mc{$\bmax^C J$} & \mc{$C_{\text{max}} / \kB N$} & \mc{$\bmin^{Q_2}
                                                                                                                J$} & \mc{$\qmin / J^2$} & \mc{$\beqh^{D} J$} & \mc{$\Pmax^D/\Pmin^D$}
      & \mc{$\beqh^{E} J$} & \mc{$\Pmax^E/\Pmin^E$} \\
      \hline
      8  & 9.902(4)   & 20.10(4) & 9.834(5)   & 1.904(3) & 8.97(1)    & 0.230(2) & 9.906(4)   & 1.40(4) &            &         \\
      12 & 10.26(1)   & 51(1)    & 10.21(1)   & 2.83(3)  & 9.72(4)    & 0.297(4) & 10.29(1)   & 1.7(1)  &            &         \\  
      16 & 10.42(1)   & 75(1)    & 10.246(3)  & 2.99(2)  & 9.76(2)    & 0.293(3) & 10.39(1)   & 1.5(1)  &            &         \\  
      20 & 10.205(2)  & 111(1)   & 10.208(1)  & 3.53(2)  & 9.98(1)    & 0.272(3) & 10.26(1)   & 1.20(3) &            &         \\  
      24 & 10.192(1)  & 190(2)   & 10.199(1)  & 4.44(3)  & 10.059(3)  & 0.244(3) & 10.199(3)  & 1.14(2) &            &         \\  
      28 & 10.180(1)  & 310(3)   & 10.188(1)  & 5.6(1)   & 10.104(2)  & 0.17(1)  & 10.176(1)  & 1.0(2)  &            &         \\  
      32 & 10.177(1)  & 457(4)   & 10.183(1)  & 6.99(5)  & 10.123(1)  & 0.11(1)  & 10.172(1)  & 1.17(4) &            &         \\  
      36 & 10.176(1)  & 662(5)   & 10.180(1)  & 8.8(1)   & 10.139(1)  & 0.05(1)  & 10.173(1)  & 1.48(5) & 10.177(1)  & 1.0(1)  \\  
      40 & 10.173(1)  & 916(8)   & 10.176(1)  & 10.8(1)  & 10.147(1)  & -0.05(1) & 10.1719(5) & 1.8(1)  & 10.175(1)  & 1.23(2) \\  
      44 & 10.1724(2) & 1237(10) & 10.1744(3) & 13.3(1)  & 10.1521(4) & -0.17(1) & 10.1716(2) & 2.1(1)  & 10.1740(2) & 1.47(3) \\  
      48 & 10.1728(4) & 1688(26) & 10.1742(4) & 16.7(2)  & 10.157(1)  & -0.35(2) & 10.1727(4) & 3.1(2)  & 10.1744(4) & 1.9(1)  \\  
    \end{tabular}
  \end{ruledtabular}
\end{table*}

\subsection{\label{sec:transition-point}Transition temperature}

\begin{figure}[htp!]
  \includegraphics[]{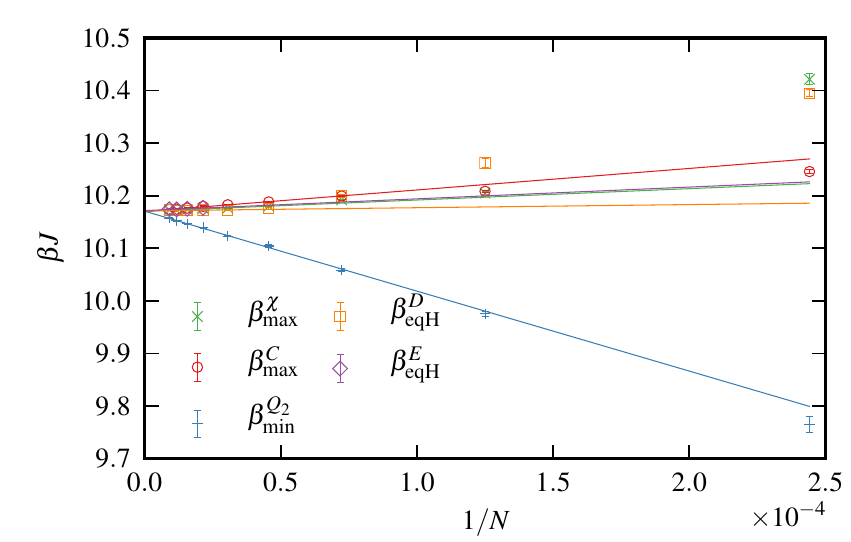}
  \caption{\label{fig:scalingbeta0} (Color online)  Finite-size
    scaling of inverse pseudotransition temperatures from
    Table~\ref{tab:pseudotemps} for $L \ge 16$ together with the best
    fits from Table~\protect\ref{tab:fittemps}, which allow to
    extrapolate the infinite-volume transition point $\beta_0$.\\[+1mm]}
\end{figure}

With $\bmax^C(L)$, $\bmax^{\chi}(L)$, $\bmin^{Q_2}(L)$, $\beqh^{D}(L)$
and $\beqh^{E}(L)$ there are various possible definitions of a
lattice-size dependent inverse pseudotransition temperature
$\beta^{*}(L)$.  For a discussion of the canonical finite-size scaling
at a first-order transition see Ref.~\onlinecite{JankeFirstOrder} and
references therein.  The inverse pseudotransition temperatures are
expected to have a displacement from the true infinite-volume
transition point $\beta_0$ which to leading order scales
proportionally to the reciprocal system size $1/L^3$:
\begin{align}
  \label{eq:10}
  \beta^{*}(L) = \beta_0 + \frac{c^{*}}{L^3} + \dotsb.
\end{align}
We test this scaling relation for all definitions of $\beta^{*}(L)$
given above by performing least-squares fits of Eq.~(\ref{eq:10}) to
the $\beta^{*}(L)$ for various ranges of lattice sizes.  The results
are given in Table~\ref{tab:fittemps}.  Fits of good quality can be
made based on all possible definitions with the limitation that we
only have very few data points for the histogram-based temperature
definitions, where the regular behavior sets in at large lattice
sizes.  The different estimates of the inverse transition temperature
$\beta_0$ and their statistical uncertainties are in good agreement
with each other.  This supports the proposed first-order nature of the
transition.  The best result is found from the $\bmax^C(L)$ data,
which yields
\begin{align}
  \label{eq:11}
  \beta_0 = 10.1700(3) / J
\end{align}
for $L \ge 24$ with $\cdof=1.12$.  This corresponds to a transition
temperature
\begin{align}
  \label{eq:12}
  T_0 = 0.098328(3) J / \kB.
\end{align}
The scaling is also visualized in Fig.~\ref{fig:scalingbeta0}.  While
it is possible to consider additional terms with higher powers of
$1/L^3$ or exponential corrections\cite{JankeFirstOrder} in the
scaling law~\eqref{eq:10}, this also leads to a higher number of free
parameters and in this case does not improve the quality of the fits.

We note that with periodic boundary conditions it may occur that the
exponential degeneracy of ground states survives partially also at
low, but finite temperatures, leading effectively to a macroscopic
degeneracy of distinct ordered states separated from each other by
free-energy barriers.  This can be understood as a number of ordered
phases $q$ that is not constant, but grows exponentially as a function
of the system size.  It has recently been
understood\cite{MuellerJankeJohnston,MuellerJohnstonJanke} that in
such a case a modified scaling law
\begin{align}
  \label{eq:15}
  \beta^{*}(L) = \beta_0 + \frac{c^{*} \ln q}{L^3} + \dotsb
\end{align}
needs to be applied, which predicts a transmuted leading system-size
dependence.  An advantage of our choice of screw-periodic boundary
conditions is that such degeneracies are mostly lifted.  In contrast
to the gonihedric plaquette model studied in
Refs.~\onlinecite{MuellerJankeJohnston,MuellerJohnstonJanke} we do not
know about any rigorous calculations of this $T>0$ degeneracy for the
3D compass model with periodic boundary conditions, but assuming a
degeneracy $\ln q \propto L^2$ the displacement of $\beta^{*}(L)$ from
the true transition point $\beta_0$ would be proportional to $1/L$
rather than to $1/L^3$.  Due to very strong finite-size effects we
cannot give a full discussion of the asymptotic scaling behavior with
periodic boundary conditions at this point.  Our (less extensive) data
for this case is compatible with the modified ansatz, but does not
allow to discriminate between the two options.  We have also checked
modified scaling relations corresponding to $\ln q \propto L^2$ and
$\ln q \propto L$ for the case of screw-periodic boundary conditions
and have found here no compelling numerical evidence against the
conventional $1/L^3$ law as reported above.

\begin{table*}[tbp]
  \caption{\label{tab:fittemps} Results of least-squares fits of the
    inverse pseudotransition temperatures $\beta^{*}(L)$
    taken from Table~\ref{tab:pseudotemps}  to estimate the 
    infinite-volume transition point $\beta_0$ by relations of the 
    form $\beta^*(L) = \beta_0 + c^{*}/L^3$.  Here $n$ is the
    number of included data points ranging from the smallest
    considered lattice size $L_{\text{min}}$ up to the largest
    $L_{\text{max}} = 48$.  $\cdof = {\chi^2}/({n - 2})$ is a
    measure to help with the estimation of the validity of the fit.
    The best fits are marked bold for each type of pseudotransition
    temperature.}
  \begin{ruledtabular}
    \begin{tabular}{
      D{.}{.}{1}
      D{.}{.}{1}
      D{.}{.}{2.5}
      D{.}{.}{3.2}
      D{.}{.}{2.6}
      D{.}{.}{3.2}
      D{.}{.}{2.5}
      D{.}{.}{3.2}
      D{.}{.}{2.5}
      D{.}{.}{3.2}
      D{.}{.}{2.5}
      D{.}{.}{1.2}
      }
      \mc{$L_{\text{min}}$} & \mc{$n$} & \mc{$\beta^{\chi}_{\text{max}, 0} J$} &
                                                                              \mc{$\cdof$} & \mc{$\beta_{\text{max,0}}^{C} J$} & \mc{$\cdof$} &
                                                                                                                                              \mc{$\beta_{\text{min}, 0}^{Q_2} J$} & \mc{$\cdof$} & \mc{$\beta_{\text{eqH}, 0}^{D}
                                                                                                                                                                                                 J$} & \mc{$\cdof$} & \mc{$\beta_{\text{eqH},0}^{E} J$} & \mc{$\cdof$} \\
      \hline
      8        & 11      & 10.176(3)             & 268.62          & 10.180(4)             & 445.25          & 10.157(4)            & 148.63          & 10.174(3)            & 259.64          &                      &                 \\
      12       & 10      & 10.170(2)             & 51.07           & 10.173(2)             & 62.05           & 10.170(2)            & 14.69           & 10.168(2)            & 86.60           &                      &                 \\
      16       & 9       & 10.169(2)             & 42.43           & 10.171(1)             & 6.85            & 10.171(1)            & 1.97            & 10.165(3)            & 78.28           &                      &                 \\
      {\bf 20} & {\bf 8} & 10.1693(5)            & 3.52            & 10.171(1)             & 7.15            & \mcb{2.5}{10.171(1)} & \mcb{3.2}{1.57} & 10.169(2)            & 22.40           &                      &                 \\   
      {\bf 24} & {\bf 7} & 10.169(1)             & 4.18            & \mcb{2.6}{10.1700(3)} & \mcb{3.2}{1.12} & 10.170(1)            & 1.57            & 10.170(1)            & 12.28           &                      &                 \\
      {\bf 28} & {\bf 6} & \mcb{2.5}{10.1702(5)} & \mcb{3.2}{1.60} & 10.1699(5)            & 1.38            & 10.170(1)            & 1.59            & \mcb{2.5}{10.171(1)} & \mcb{3.2}{2.85} &                      &                 \\
      32       & 5       & 10.170(1)             & 2.10            & 10.170(1)             & 1.64            & 10.170(1)            & 1.79            & 10.172(1)            & 2.64            &                      &                 \\
      {\bf 36} & {\bf 4} & 10.170(1)             & 3.11            & 10.169(1)             & 2.34            & 10.170(1)            & 2.02            & 10.172(1)            & 3.95            & \mcb{2.5}{10.172(1)} & \mcb{1.2}{2.84} \\
      40       & 3       & 10.172(2)             & 2.66            & 10.171(2)             & 2.16            & 10.171(2)            & 1.80            & 10.173(2)            & 4.51            & 10.173(2)            & 3.22            \\
    \end{tabular}
  \end{ruledtabular}
\end{table*}

\begin{figure}[htp!]
  \includegraphics[]{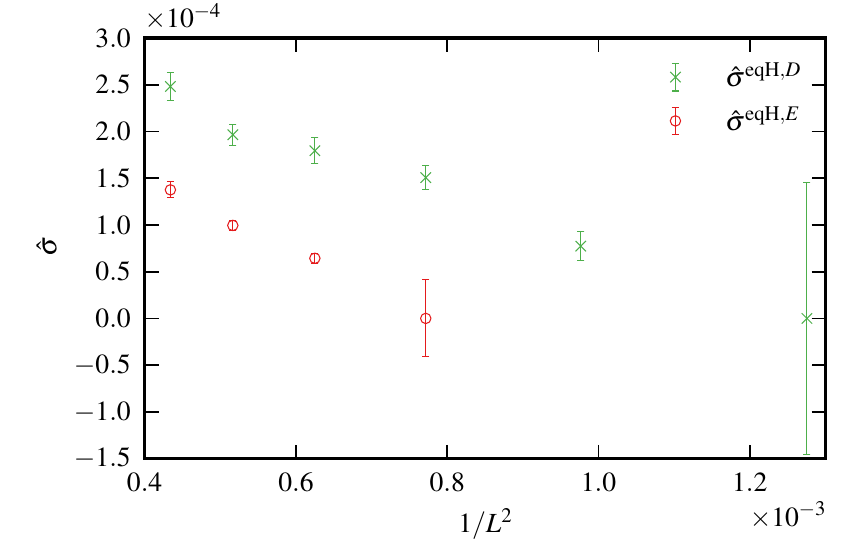}
  \caption{\label{fig:ift} (Color online)  Reduced interface tensions
    $\hat{\sigma}(L)$ calculated from $P(D)$ histograms at
    $\beqh^D(L)$ and from $P(E)$ histograms at $\beqh^E(L)$ plotted
    over $1/L^2$ for $L \ge 28$.}
\end{figure}

\subsection{\label{sec:interface-tension}Interface tension}

On lattices of size $L^3$ the suppression of the minimum between the
two peaks of the probability distribution of the energy or the order
parameter at a first-order phase transition is expected to grow
exponentially with $L^2$:
\begin{align}
  \label{eq:13}
  \Pmax(L)/\Pmin(L) \propto \e^{2 \beta \sigma L^2}.
\end{align} 
Configurations corresponding to $\Pmin(L)$ are in a mixture of the
ordered and the disordered phases with interfaces that contribute an
excess free energy of $2 \sigma L^2$, where the free-energy density
$\sigma$ is the interface tension.\cite{JankeFirstOrder}  We compute
lattice size dependent estimates of the reduced interface tension
$\hat\sigma(L) = \beta\sigma(L)$ from the double-peaked probability
distributions $P(D)$ at $\beqh^D(L)$ or $P(E)$ at $\beqh^E(L)$ with
the relation
\begin{align}
  \label{eq:14}
  \hat\sigma(L) = \frac{1}{2 L^2}
  \ln\left[\frac{\Pmax(L)}{\Pmin(L)}\right],
\end{align}
where $\Pmax(L)/\Pmin(L)$ is the ratio of the estimated probabilities
in the peak and in the dip as taken from Table~\ref{tab:pseudotemps}.
In Fig.~\ref{fig:ift} the results are plotted over $1/L^2$ for $L \ge
28$, which excludes the irregular behavior for the small lattices.
While the reduced interface tension does not yet reach its asymptotic
constant value on the lattice sizes studied here, $\hat\sigma(L)$
grows with $L$ and does not appear to vanish in the limit of large
systems, which otherwise would be an argument against the first-order
nature of the transition.  From the available data an approximate
infinite-volume value of $\hat\sigma \approx 3 \times 10^{-4}$ can be
anticipated.

\section{\label{sec:summary}Summary and conclusions}

In this paper we have presented an extensive Monte Carlo investigation
of the classical compass model on the simple-cubic lattice.  Our
results show that directional ordering is present in a low-temperature
phase, which is reached via a thermal first-order transition from a
disordered high-temperature phase.  By a detailed finite-size scaling
analysis we could determine a precise estimate of the transition
temperature $T_0 = 0.098328(3) J / \kB$.  This value agrees with the
one mentioned in an earlier publication,\cite{Wenzel2011a} but the
high-temperature series expansions presented in
Ref.~\onlinecite{Oitmaa2011} could not identify this phase transition.
First-order transitions are generally difficult to detect by these
techniques, in particular when no low-temperature series are
available. 

The recently discovered (and for the gonihedric plaquette model
numerically confirmed) influence of a macroscopic degeneracy of the
low-temperature phases on the leading finite-size scaling behavior of
first-order phase
transitions\cite{MuellerJankeJohnston,MuellerJohnstonJanke} renews the
interest in a precise characterization of the ground-state and
low-temperature degeneracies of the compass model.  A rigorous
treatment along the lines of
Refs.~\onlinecite{Nussinov2004,PietigWegner1,PietigWegner2} for the
closely related 120$^\circ$ model and the gonihedric model is beyond
the scope of the present paper focusing on an accurate determination
of the first-order character of the phase transition, but would
certainly be a worthwhile project for future studies, especially with
a view on the ``order-by-disorder'' mechanism.

Due to the negative-sign problem, quantum Monte Carlo
simulations of the 3D compass model are out of reach.  However, while
additional quantum fluctuations may destroy directional ordering at
low temperatures in the quantum model, from Ginzburg-Landau theory one
generally expects the nature of the phase transition to be the same in
the quantum model as in the classical model.  Symmetry considerations
for the nematic-like type of order parameter of the $t_{2g}$ compass
model support the expectation of a continuous transition in 2D and a
first-order transition in 3D, just as observed in the Monte Carlo
simulations.  Taken together, we firmly anticipate a first-order phase
transition to occur also in the quantum compass model and look forward
to experimental studies of directional ordering in non-low dimensional
samples.

\begin{acknowledgments}
  We thank C. Hamer, J. Oitmaa and W. Selke for useful discussions
  initiating this work, as well as A. Rosch and S. Trebst for further
  helpful conversations.  Partial support by the Deutsche
  Forschungsgemeinschaft (DFG) through Graduate School GSC185
  ``BuildMoNa'' and the Deutsch-Franz\"osische Hochschule (DFH-UFA)
  through the binational German-French Graduate School under grant
  number CDFA-02-07 is gratefully acknowledged.  M.G. thanks the
  Bonn-Cologne Graduate School of Physics and Astronomy (BCGS) for
  support.
\end{acknowledgments}

\end{document}